\documentclass[epj]{svjour}
\usepackage{graphics}
\usepackage{epstopdf}
\usepackage{amssymb}
\usepackage{bm}
\usepackage{amsmath}
\usepackage{nicefrac}
\usepackage{url}
\begin{document}
\title{
  Di--nucleon structures in homogeneous nuclear matter 
  based on two- and three-nucleon interactions}
\author{
  Hugo F. Arellano\inst{1,2} \and 
  Felipe Isaule\inst{1} \and
  Arnau Rios\inst{3}
}                     
\offprints{}          
\institute{
Department of Physics - FCFM, University of Chile,
Av. Blanco Encalada 2008, Santiago, Chile
\and
CEA, DAM, DIF, F-91297 Arpajon, France
\and
Department of Physics,
Faculty of Engineering and Physical Sciences,
University of Surrey, Guildford,
Surrey GU2 7XH, United Kingdom
}
\date{Received: date / Revised version: date}
%
\abstract{
  We investigate homogeneous nuclear matter within the 
  Brueckner-Hartree-Fock (BHF) approach in the limits of
  isospin-symmetric nuclear matter (SNM) as well as pure neutron matter
  at zero temperature.
  The study is based on realistic representations of the
  internucleon interaction as given by
  Argonne v18, Paris, Nijmegen I and II potentials,
  in addition to chiral N$^{3}$LO interactions, 
  including three-nucleon forces up to N$^{2}$LO.
  Particular attention is paid to the presence of di-nucleon 
  bound states structures in 
  $^1\textrm{S}_0$ and
  $^3\textrm{SD}_1$ channels,
  whose explicit account becomes crucial for the stability of
  self-consistent solutions at low densities.
  A characterization of these solutions and associated
  bound states is discussed. 
  We confirm that coexisting BHF single-particle solutions in SNM,
  at Fermi momenta in the range $0.13-0.3$~fm$^{-1}$,
  is a robust feature under the choice of realistic internucleon potentials.
\PACS{
      {21.65.-f}{Nuclear matter}\and
      {21.45.Bc}{Two-nucleon system}\and
      {21.65.Mn}{Equations of state of nuclear matter}
}
} 
\titlerunning{Di-nucleon structures in homogeneous nuclear matter...}
\authorrunning{Arellano \emph{et. al.}}
\maketitle
\section{Introduction}
\label{intro}
One of the main goals in theoretical nuclear physics is that
of accounting for nuclear structures and processes starting from
the basic interaction among their constituents. 
If sub-hadronic degrees of freedom (i.e. quarks and gluons) 
are not treated explicitly, then this goal relies on realistic
representation of the bare internucleon interaction.
Such is the case of modern interactions based on quantum-field 
models, where strengths and form factors are adjusted to best 
reproduce nucleon-nucleon (\emph{NN}) scattering observables 
as well as properties of the deuteron, 
the only \emph{NN} bound state in free space.
The inclusion of three-nucleon (3N) forces become subject to
constraints from three-body bound state data and/or
homogeneous nuclear matter~\cite{Baldo99,Entem2003}.

Homogeneous nuclear matter, a hypothetical infinite medium of neutrons
and protons, is among the simplest many-body nuclear systems.
In principle, all properties of this system should be inferred 
from the bare interaction among its constituents.
In this context the Brueckner-Hartree-Fock (BHF) non-relativistic 
approach at zero temperature offers a well defined framework which enables
the evaluation of the energy of the system as a function of the
nucleon density $\rho=\rho_p+\rho_n$, 
and isospin asymmetry $\beta=(\rho_n-\rho_p)/(\rho_n+\rho_p)$,
with $\rho_p$ ($\rho_n$) denoting proton (neutron) 
density~\cite{Baldo99,Dickhoff2008}.
Extensive applications of the BHF approach has served to assess the 
consistency of the model to account for saturation properties of 
isospin-symmetric ($\beta=0$) 
nuclear matter~\cite{Baldo99,Dewulf03,Song98}.
Not only that but also the resulting BHF $g$ matrix at positive 
energies has served as an important tool to construct \emph{NN} 
effective interactions, subsequently used in the evaluation of 
microscopic optical model potentials for nucleon scattering off 
finite nuclei~\cite{Amos00}.

In this work we investigate homogeneous nuclear matter in the
framework of the BHF approach at zero temperature considering 
realistic representations of the bare \emph{NN} interactions. 
Particular attention is paid to the manifestation of two-nucleon
bound state structures (di-nucleons), expressed as singularities
in the $g$ matrix in the search process for self-consistent 
solutions of single-particle (sp) spectra.
As such, this work represents an extension of the investigation 
reported in Ref.~\cite{Arellano15} based on Argonne $v_{18}$ 
(AV18)~\cite{Wir95}, where an explicit account 
for di-nucleon structures in symmetric nuclear matter was first 
reported.
Among the main findings reported in that work we mention:
\textsl{a)} 
Nucleon effective masses at low densities can 
reach up to four times the bare nucleon mass;
\textsl{b)} 
Large size di-nucleon bound states take 
place at sub-saturation densities; and
\textsl{c)} 
Coexisting sp spectra are identified at low densities,
that is to say two distinct sp fields meet self-consistency
at a same density.
Here we investigate the robustness of these features in SNM
under the choice of the bare \emph{NN} interaction,
in addition to their manifestation in the extreme case of 
pure neutron matter.

The BHF approach for interacting nucleons in nuclear matter
can be thought as the lowest-order approximation of
Brueckner-Bethe-Goldstone (BBG) theory or Self-Consistent
Green's Function theory at zero temperature \cite{Dickhoff2008}.
The former is based on the hole-line expansion for the ground 
state energy \cite{Baldo99}, where Goldstone diagrams are grouped 
according to their number of hole lines, with each group summed 
up separately.
The BHF approximation results from the summation of the two-hole-line 
diagrams, with the \emph{in-medium} two-body scattering matrix 
calculated self-consistently with the sp energy spectrum.
Although a sp potential is introduced as an auxiliary quantity,
its choice conditions the rate of convergence of the expansion for
the binding energy.
Studies reported in Ref.~\cite{Song98} lead to conclude that the 
continuous choice for the auxiliary potential yields better convergence
over the so called \emph{standard choice}, where the sp potential
is set to zero above the Fermi energy.
Thus, we base this work on the continuous choice for the sp potentials.

This article is organized as follows.
In Sec.~\ref{framework} we layout the theoretical framework upon
which we base the study of homogeneous nuclear matter at zero temperature.
In Sec.~\ref{results} we present results for symmetric nuclear matter
as well as neutronic matter, discuss associated effective masses
and occurring \emph{in-medium} di-nucleon structures.
Additionally, we discuss extent to which 
Hugenholtz-van Hove theorem~\cite{Hugenholtz1958} for sp energies is met.
In Sec.~\ref{summary} we present a summary and the main 
conclusions of this work.

\section{Framework}
\label{framework}
In BBG theory for homogeneous nuclear matter the $g$ matrix depends
on the density of the medium, characterized by the Fermi momentum $k_F$,
and a starting energy $\omega$.
To lowest order in the BHF approximation for nuclear matter
in normal state,
when only two-body correlations are taken into account,
the Brueckner $G$ matrix satisfies
\begin{equation}
\label{bbg0}
G(\omega)=v+v\,\frac{Q}{\omega+i\eta-\hat h_1-\hat h_2}\,G(\omega)\,,
\end{equation}
with $v$ the bare interaction between nucleons,
$\hat h_{i}$ the sp energy of nucleon $i$ ($i=1,2$),
and $Q$ the Pauli blocking operator which for nuclear matter
in normal state takes the form
\[
Q|\boldsymbol{ p\, k}\rangle =
\Theta(p-k_F)\Theta(k-k_F) |\boldsymbol{ p\, k}\rangle \;.
\]
The solution to Eq.~(\ref{bbg0}) enables the evaluation of
the mass operator
\begin{equation}
\label{mass}
M(k;E)=\sum_{\mid\boldsymbol p\mid\leq k_F}
\langle \textstyle{\frac{1}{2}}(\boldsymbol k-\boldsymbol p)
| g_{\boldsymbol K}(E+e_p) |
\textstyle{\frac{1}{2}}(\boldsymbol k-\boldsymbol p)\rangle\;,
\end{equation}
where the $g$ matrix relates to $G$ through
\begin{equation}
  \label{Gg}
 \langle 
 {\boldsymbol k'} {\boldsymbol p'} 
 | G(\omega) |
 {\boldsymbol k} {\boldsymbol p}
 \rangle
=
 \delta({\boldsymbol K'}-{\boldsymbol K})
  \langle 
  \textstyle{\frac12}( {\boldsymbol k'}-{\boldsymbol p'})|
  g_K(\omega) |
  \textstyle{\frac12}({\boldsymbol k}-{\boldsymbol p})\rangle \;.
\end{equation}
Here  $\boldsymbol K$  ($\boldsymbol K')$ 
denotes the total momentum of the \emph{NN} pair before (after) interaction,
with $\boldsymbol K = \boldsymbol k+\boldsymbol p$, and
$\boldsymbol K' = \boldsymbol k'+\boldsymbol p'$,
so that the Dirac delta functions expresses the momentum conservation.
The sp energy becomes defined in terms of an auxiliary field $U$,
\begin{equation}
\label{esp}
e(p)=\frac{p^2}{2m} + U(p)\,,
\end{equation}
with $m$ the nucleon mass taken as the average of proton and neutron masses.
In the BHF approximation the sp potential is given by the
on-shell mass operator,
\begin{equation}
\label{usp}
U(k)=\textsf{Re}\, M[k;e(k)]\;,
\end{equation}
self-consistency requirement which can be achieved iteratively.
We have used the continuous choice for the sp fields, 
so that this condition is imposed at all momenta $k$ \cite{Baldo00}.

\section{Results}
\label{results}
We have proceeded to obtain self-consistent solutions for
the sp fields $U(k)$ in infinite nuclear matter 
at various densities, specified by
Fermi momenta $k_F\lesssim 2.5$~fm$^{-1}$.
These searches comprise isospin-symmetric nuclear matter
and pure neutron matter.
The internucleon interactions considered in this study are 
AV18~\cite{Wir95}, 
Paris~\cite{Paris}, 
Nijmegen I and II bare potentials~\cite{Nijmegen}.
In addition to these potentials we include a chiral
effective-field-theory ($\chi$EFT) interaction based 
on chiral perturbation theory.
The resulting bare interaction is constructed with nucleons and
pions as degrees of freedom, with the two-nucleon (2N) part fit 
to \emph{NN} data. 
We consider the chiral 2N force (2NF) up to 
next-to-next-to-next-to-leading order (N$^{3}$LO) given
by Entem and Machleidt \cite{Entem2003}.
We also consider chiral 3N forces (3NF) in
N$^{2}$LO, using a density-dependent 2NF at the two-body level
\cite{Holt2010,Hebeler2010a}. 
This density-dependent contribution does not contain correlation 
effects \cite{Carbone2014}, and is added to the bare 2NF
in the calculation of the $G$ matrix. 
The corresponding Hartree-Fock contribution is subtracted 
at each iteration to avoid any double counting.
For this chiral 3NF contribution,
we use the low energy constants $c_D = -1.11$ and $c_E = -0.66$,
reported in Ref. \cite{Nogga2006}, which describe the $^3$H and
$^4$He binding energies with unevolved \emph{NN} interactions.

All applications include partial waves up to $J=7$
in the \emph{NN} total angular momentum.
For the numerical methodology to treat di-nucleons during 
self-consistency search
we refer the reader to Ref. \cite{Arellano15}.
Files containing self-consistent solutions for sp potentials 
can be retreived from Ref.~\cite{omponline}.

\subsection{Symmetric nuclear matter}
\label{snm}

The study of SNM requires to take into account \emph{NN} 
states with total isospin $T=0$, and $T=1$. 
As a result, one has to include the attractive
${}^3\textrm{SD}_{1}$, 
${}^3\textrm{PF}_{2}$
and ${}^1\textrm{S}_{0}$ channels.
As reported in Ref.~\cite{Arellano15}, the calculation
of the on-shell mass operator to obtain $U(k)$ requires
the evaluation of $g_{K}(\omega)$ at various configurations of
total momentum $K$ of the \emph{NN} pair and starting energy
$\omega=e(p)+e(k)$. 
In the process the $g$ matrix is sampled over regions where
it becomes singular, near or at the occurrence of 
\emph{in-medium} bound states in these channels. 
This feature, investigated in the context of AV18 interaction, 
has led to unveil coexisting sp solutions at Fermi momenta 
slightly below 0.3~fm$^{-1}$,
that is to say different solutions that meet self-consistency 
at the same $k_F$.
Details about how these coexisting solutions are disclosed
are given in the same reference.
In this work we proceed in the same way.

Once a sp solution $U(k)$ is obtained for a given Fermi
momentum we can evaluate the energy per nucleon $E/A$,
which in the case of two-body forces is given by

\begin{equation}
  \label{boa}
  \frac{E_{2N}}{A} = 
  \frac{\sum_{k}
    n(k) 
    \left [
      \frac{k^2}{2m} + 
      \textstyle{\frac12}U(k) 
  \right]}
  {\sum_{k} n(k)}\;.
\end{equation}
In this work we use $n(k) = \Theta(k_F-k)$,
i.e. nuclear matter in normal state.
When 3NFs are included in BHF calculations, these enter at two levels.
First, a density-dependent effective two-body interaction is added to the 
bare 2NF in a standard $G$-matrix calculation. 
In addition, the total energy has to be corrected to avoid 
double counting of the 3NF contribution \cite{Hebeler2010a,Carbone2013}. 
At the lowest order this can be achieved by subtracting the 
Hartree-Fock contribution due to 3NFs only:
\begin{align}
\frac{E_{3N}}{A}=\frac{E_{2N}}{A} &-\frac{1}{12}\frac{3}{k_F^3}\int_0^{k_F} k^2 dk \, \Sigma_{HF}^{3NF}(k) \, .
\end{align}
We stress that the Hartree-Fock self-energy $\Sigma^{3NF}_{HF}$ coming 
from the 3N force is calculated from an effective 2N potential at the 
lowest order, in keeping with the procedure established 
Ref.~\cite{Hebeler2010a}.

\begin{figure} [ht]  
\resizebox{0.50\textwidth}{!}{%
  \includegraphics{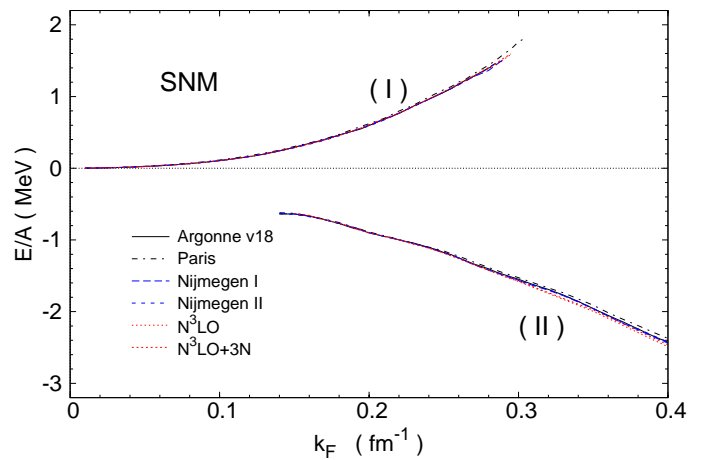}
}
\caption{
  Energy per nucleon for isospin-symmetric nuclear matter
  as function of Fermi momentum $k_F$.
  Solid, long-, medium- and short-dashed curves correspond to
  AV18, Paris, Nijmegen I and II, respectively. 
  Dotted and dash-dotted curves represent solutions for 
  N$^3$LO and N$^3$LO+3N chiral interactions, respectively.
}
\label{boa_snm_i}       
\end{figure}

In Fig.~\ref{boa_snm_i} we present results for the energy per 
nucleon $E/A$ as function of $k_F$ for symmetric nuclear matter.
Here, 
  solid, long-, medium- and short-dashed curves denote 
  AV18, Paris, Nijmegen I and II solutions, respectively. 
  Dotted and dash-dotted curves represent solutions based on
  N$^3$LO and N$^3$LO+3N chiral interactions, respectively.
Labels I and II are used to distinguish the two families of
solutions. 
We are aware that, from a physical point of view,
the energy of the system should be uni-valuated. 
The purpose of this figure in displaying separately $E/A$
for the two phases is that of providing a global characterization 
of the sp solutions. 
An actual evaluation of the energy of the system at a given 
$k_F$ would require a more comprehensive analysis, 
considering contributions from di-nucleons in the different channels
and competing phases I and II.
Under such considerations the scope of the BHF approximation 
would become limited.

The results presented in Fig.~\ref{boa_snm_i} show that
all interactions considered yield nearly
identical behavior in the range $0\leq k_F\leq 0.4$~fm$^{-1}$.
Additionally, they all exhibit coexisting sp solutions at $k_F$
in the range between ${\sim}$0.13 and $0.28-0.30$~fm$^{-1}$.
In this regard the feature of coexistence is robust under
the bare internucleon bare interaction.

\begin{figure} [ht]  
  \resizebox{0.50\textwidth}{!}{ \includegraphics{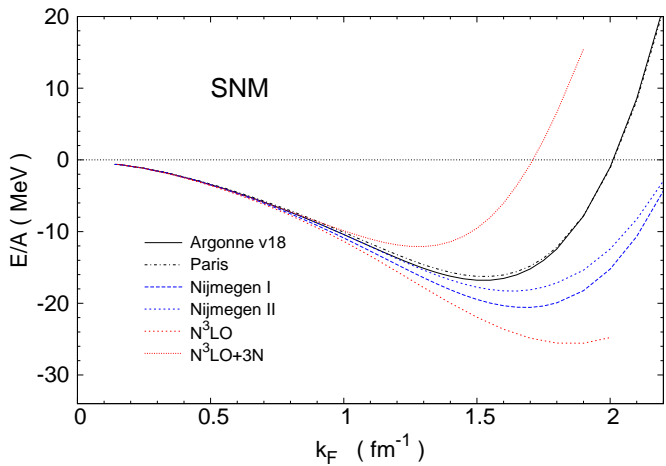}}
\caption{
  Energy per nucleon for isospin-symmetric nuclear matter
  as function of Fermi momentum $k_F$.
  Curves follow the same convention as in Fig.~\ref{boa_snm_i}.
}
\label{boa_snm_ii}       
\end{figure}

In Fig.~\ref{boa_snm_ii} we present results for the
energy per nucleon $E/A$ in SNM as a function of $k_F$ 
for solutions in phase II.
We use the same convention of curve patterns as in Fig.~\ref{boa_snm_i}.
In this case the interactions exhibit different behaviors,
resulting in different saturation points.
As observed, chiral interactions are the ones which yield
extreme values for the density and binding energy at saturation.
On the one side 
N$^{3}$LO saturates  at $k_F=1.85$~fm$^{-1}$, with $E/A=-25.7$~MeV, whereas
N$^{3}$LO+3N does so at $k_F=1.30$~fm$^{-1}$, with $E/A=-12.1$~MeV.
The former becomes much too bound at a density $\sim$$2.7\rho_0$,
with $\rho_0=0.16$~fm$^{-3}$, the accepted saturation density.
The latter, instead, saturates near the correct density but becomes
underbound by $\sim$4~MeV relative to the accepted value of
$16\pm 1$~MeV.
These results are comparable to those reported 
in Ref.~\cite{Carbone2014}.
Furthermore, at $k_F$ below $\sim\!1$~fm$^{-1}$, 
i.e. matter density below $0.07$~fm$^{-3}$, 
the behavior of $E/A$ appears insensitive to the interaction.
This feature is in agreement with recent reports based on 
BHF and Monte Carlo calculations 
using chiral interactions~\cite{Tews2016,Sammarruca2015}.

Another feature we note from Fig.~\ref{boa_snm_ii} 
is the similarity between AV18 and Paris potentials 
in their $E/A$ vs $k_F$ behavior.
Their resulting saturation energies are $-16.8$ and $-16.3$~MeV,
respectively, with both interactions saturating at a density near
$1.4\rho_0$.
In the cases of Nijmegen I and II saturation occurs
at $2\rho_0$ and $1.8\rho_0$, respectively, while their respective
binding occur at $-20.6$ and $-18.3$~MeV.

The results we provide here are in reasonable agreement with 
those reported elsewhere~\cite{Lombardo2006}.
What is new in these results is the actual account for di-nucleon
singularities in the $g$ matrix to obtain self-consistent sp fields
within BHF. 
Unfortunately there is now way to artificially suppress 
di-nucleon occurrences, without altering the bare interaction,
in order to isolate the role of \emph{in-medium} bound states.

\begin{figure} [ht]  
\resizebox{0.50\textwidth}{!}{%
  \includegraphics{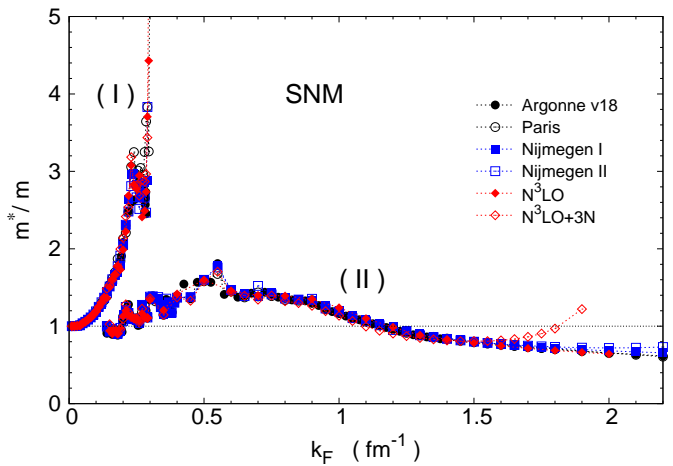}
}
\caption{
  Nucleon effective mass in isospin-symmetric nuclear
  matter as function of Fermi momentum $k_F$.
Filled and empty circles represent solution for
AV18 and Paris potentials, respectively.
Filled and empty squares correspond to Nijmegen I and II potentials, 
respectively.
Filled and empty diamonds denote solutions based on 
N$^3$LO and N$^3$LO+3N chiral interactions, respectively.
Dotted lines are used to guide the eye.
}
\label{meff_snm}       
\end{figure}

The study of nucleon effective masses $m^{*}$ has been subject of 
interest in various sub-fields~\cite{Chamel2013,Baldo2014}.
The calculated sp spectra of Eq.~(\ref{esp}) 
allows us to evaluate the effective mass
\begin{equation}
\label{masseff}
\frac{m^*}{m} = 
\frac{k_F}{m}\left [
\frac{\partial e(k)}{\partial k} \right ]^{-1}_{k=k_F}\,,
\end{equation}
with $m$ the nucleon mass.
In Fig.~\ref{meff_snm} we plot the calculated
effective-to-bare mass ratio $m^*/m$ as a function of
Fermi momentum based on the six interactions we have discussed.
Filled and empty circles correspond to results for
AV18 and Paris potentials, respectively.
Filled and empty squares correspond to Nijmegen I and II potentials,
respectively.
Filled and empty diamonds denote solutions based on
N$^3$LO and N$^3$LO+3N chiral interactions, respectively.
Labels I and II refer to solutions in phase I and II, respectively.

A peculiar feature observed in Fig.~\ref{meff_snm} is the
occurrence of $m^{*}/m>1$ at Fermi momenta below $\sim$1~fm$^{-1}$.
In the case of phase I, which starts at $k_F=0$, 
the effective mass grows from the bare mass $m$ up to $\sim\!4m$ 
near the maximum $k_F$ of phase I, 
consistent with findings reported in Ref.~\cite{Arellano15}.
We also note that the trend followed by $m^*/m$ vs $k_F$ is very 
similar for all the interactions considered, 
an indication of the robustness of the results under changes of
the bare internucleon potential. 
In the case of phase II, the range where $m^*>m$ is
restricted to $\sim\!0.2<k_F\lesssim 1.1$~fm$^{-1}$,
or equivalently $\sim\!0.003<\rho/\rho_0\lesssim 0.6$.
As discussed in Refs.~\cite{Arellano15,Isaule2016}, 
this feature is closely related to the occurrence 
of di-nucleon bound states.
For $k_F$ near normal densities, 
i.e. in the range $1.4-1.5$~fm$^{-1}$, 
the ratio $m^*/m$ lies within the interval $0.78-0.85$ for
all interactions,
feature consistent with the typical values of nucleon effective masses.

\subsection{Pure neutron matter}
\label{neutronic}
The case of pure neutron matter features full suppression of the
deuteron channel. Therefore, singularities of $g_K(\omega)$ 
represent \emph{in-medium} bound states formed by neutron pairs, 
i..e. di-neutrons.
The treatment of these singularities is the same as that applied
in SNM reported in Ref.~\cite{Arellano15}.
In contrast to the case of SNM, however, 
no coexisting sp spectra are found.

\begin{figure}  [ht]  
\resizebox{0.50\textwidth}{!}{\includegraphics{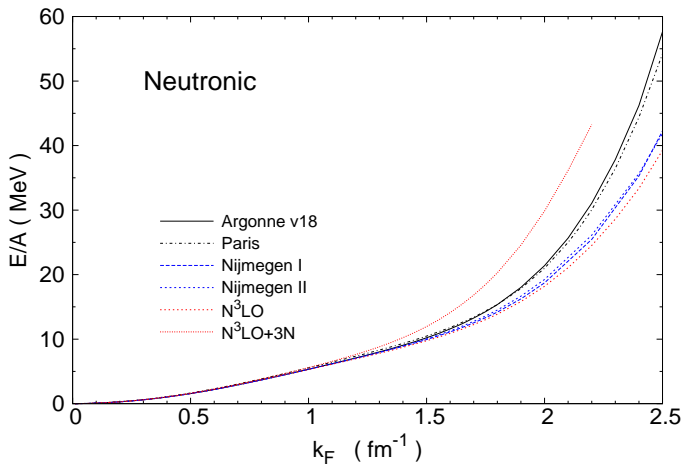} }
\caption{
  Energy per nucleon for pure neutron matter
  as function of Fermi momentum $k_F$.
  Curves follow the same convention as in Fig.~\ref{boa_snm_i}.
}
\label{boa_n}       
\end{figure}

In Fig.~\ref{boa_n} we present results for the energy 
per nucleon, $E/A$, for neutronic matter as a function of $k_F$.
We use the same convention of curve patterns as in Fig.~\ref{boa_snm_i}.
As observed, all interactions yield nearly identical
energy per nucleon up to $k_F\sim 1$~fm$^{-1}$, 
departing from each other at Fermi momenta above $1.2$~fm$^{-1}$.
All interactions yield monotonic growing $E/A$ as function
of $k_F$, with N$^{3}$LO+3N providing the highest slope.
As in the case of SNM, AV18 and Paris potentials behave very
similarly.
The smallest slope in the energy comes from N$^{3}$LO, 
although both Nijmegen I and II present similar density dependence.
For $k_F$ below $\sim\!1.2$~fm$^{-1}$, i.e. neutron densities below
$0.06$~fm$^{-3}$, all interactions exhibit nearly the same 
behavior, in agreement with other reports
based on chiral interactions~\cite{Tews2016,Sammarruca2015}.

\begin{figure} [ht]  
\resizebox{0.50\textwidth}{!}{%
  \includegraphics{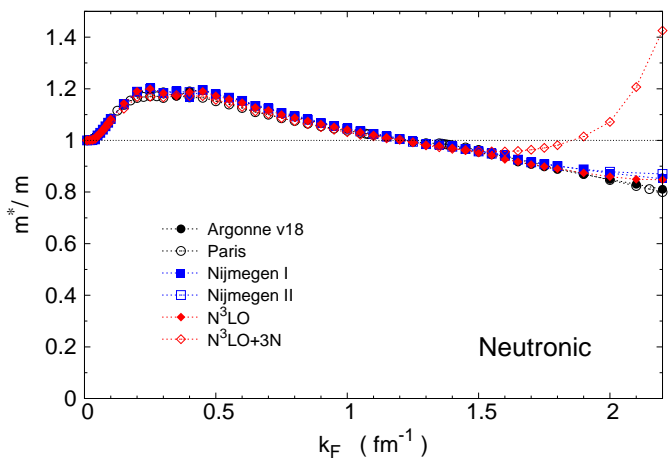}
}
\caption{
  Neutron effective mass in pure neutron matter
  as function of Fermi momentum $k_F$.
  Symbols follow the same convention as in Fig.~\ref{meff_snm}.
}
\label{meff_n}       
\end{figure}

Effective masses associated to the sp fields for pure neutron 
matter are shown in Fig.~\ref{meff_n}.
Here we consider all six interactions included in the previous 
analysis, following the same symbol convention as in 
Fig.~\ref{meff_snm}. 
As in the case of SNM, all interaction follow a very similar
behavior as function of $k_F$, with only the chiral
interaction N$^{3}$LO+3N departing from the rest at $k_F$
above 1.5~fm$^{-1}$.
It is also clear that the neutron effective mass
is greater than its bare mass at $k_F$ in the range
$0.04-1.1$~fm$^{-1}$, with a maximum value of 
$\sim\!$$1.2 m$
at $k_F$ in the range $0.25-0.5$~fm$^{-1}$.
In the case of N$^{3}$LO+3N interaction the ratio $m^{*}/m$
exhibits a growth at $k_F$ above 1.5~fm$^{-1}$, 
which could be attributed to the relevance of the 
(density-dependent) 3\emph{N} force at such high densities.
Apart from this interaction at high densities,
the behavior of effective masses featuring $m^{*}/m>1$
is also robust under the choice of bare interaction 
being considered.

\subsection{Di-nucleons within BHF}
\label{dinucleons}
As mentioned above, the occurrence of singularities in the
$g$ matrix denotes the presence of bound states. 
This feature becomes explicit with the use of the Lehmann 
spectral representation for the $g$ matrix~\cite{Dickhoff2008}
\begin{equation}
  \label{lehmann}
  g_{K}(\omega) = v + 
  \sum_{\alpha} 
  v\frac{|\alpha\rangle\langle\alpha|}{\omega+i\eta-\epsilon_\alpha}Qv\;,
\end{equation}
where $\alpha$ runs over discrete and continuous states,
$|\alpha\rangle$ is an eigenstate of the Hamiltonian
$\hat H=\hat h_1+\hat h_2 + v$, with eigenenergy $\epsilon_\alpha$.
The way to infer the energy of bound states for a given pair 
momentum $K$ is by imposing~\cite{Arellano15}
\begin{equation}
  \label{det}
\det [1-v\Lambda_K(\omega)]=0\;,
\end{equation}
with $\Lambda_K(\omega)=Q/(\omega-\hat h_1-\hat h_2)$,
the BHF particle-particle propagator.
The energy of the bound state is obtained from the difference
\begin{equation}
  \label{bstate}
  b\equiv \omega-\omega_{th}\;,
\end{equation}
where $\omega_{th}$ corresponds to the lowest (threshold) 
particle-particle energy allowed by the Pauli blocking operator.

\begin{figure} [ht] 
\resizebox{0.50\textwidth}{!}{ \includegraphics{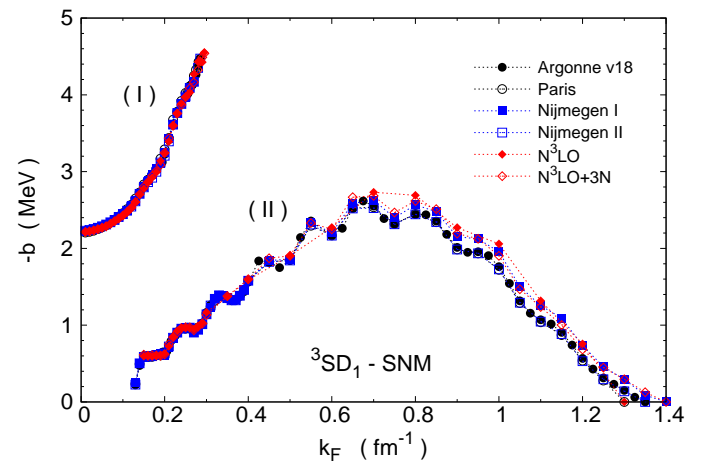} }
\caption{
  \emph{In-medium} deuteron binding energy in isospin-symmetric nuclear 
   matter as function of the Fermi momentum $k_F$.
  Symbols follow the same convention as in Fig.~\ref{meff_snm}.
}
\label{bnn3sd1snm}       
\end{figure}

We investigate the occurrence of di-nucleon bound states in SNM
in the channels $^3\textrm{SD}_1$ and $^{1}\textrm{S}_0$ as a 
function of the Fermi momentum for all six \emph{NN} interactions
under study. 
The condition given by Eq.~\ref{det} can be investigated
for selected values of $K$, the momentum of the \emph{NN} pair.
In the following we focus on center-of-mass at rest ($K=0$).
Fig.~\ref{bnn3sd1snm} shows results obtained for the 
di-nucleons in the deuteron channel, where
we follow the same symbol convention as in Fig.~\ref{meff_snm}.
Labels I and II indicate solutions for phase I and II,
respectively.
From these results we observe bound states in channel
$^{3}\textrm{SD}_{1}$ take place in phase I 
at momenta over the range $0\leq k_F\lesssim 0.3$~fm$^{-1}$,
featuring increasing binding.
The highest binding takes place at the upper edge of phase I,
where $b\approx -\!4.5$~MeV, nearly twice the binding energy of the
deuteron in free space.
It is also clear that all six interactions yield nearly the same
binding.
Phase II, in turn, shows bound states from $k_F\approx 0.14$~fm$^{-1}$
up to $k_F$ between 1.3 and 1.4~fm$^{-1}$, close to the 
accepted Fermi momentum at saturation.
The maximum binding takes place at $k_F\approx 0.5$~fm$^{-1}$, 
where $b\approx -\!2.6$~MeV.
Overall, all interactions display the same behavior for $b$.

\begin{figure}   [ht] 
\resizebox{0.50\textwidth}{!}{\includegraphics{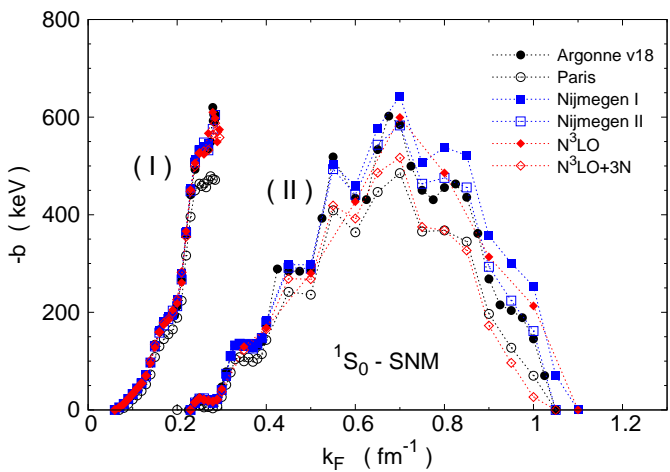}}
\caption{
  \emph{In-medium} di-nucleon binding energy in channel $^1\textsf{S}_0$ 
  for isospin-symmetric nuclear matter 
  as functions of the Fermi momentum $k_F$.
  Symbols follow the same convention as in Fig.~\ref{meff_snm}.
}
\label{bnn1s0snm}       
\end{figure}

Results for the $^{1}\textrm{S}_{0}$ channel in SNM are shown
in Fig.~\ref{bnn1s0snm} using the same notation as in the
previous figure.
Note that in this case the energy scale is expressed in keV units.
As a result, differences in the binding energy from the 
different \emph{NN} interactions appear enhanced.
With the exception of Paris potential (open circles),
the trend followed by $b$ in phase I is quite similar among the
other five interactions, leading to a maximum binding of
about 600~keV at $k_F\approx 0.28$~fm$^{-1}$.
Note also that for $k_F$ below $\sim\!0.06$~fm$^{-1}$ no
di-nucleon bound states take place, feature shown by all
six interactions. This is consistent with the fact that
no bound state takes place at zero density (free space) 
in the $^{1}\textrm{S}_{0}$ channel.
In the case of phase II, di-nucleons take place
from $k_F\gtrsim 0.2$~fm$^{-1}$ up to about $k_F$ slightly
above 1~fm$^{-1}$. 
The maximum binding takes place at $k_F\approx 0.7$~fm$^{-1}$,
with $b$ in the range $500-650$~keV.

\begin{figure} [ht] 
\resizebox{0.50\textwidth}{!}{\includegraphics{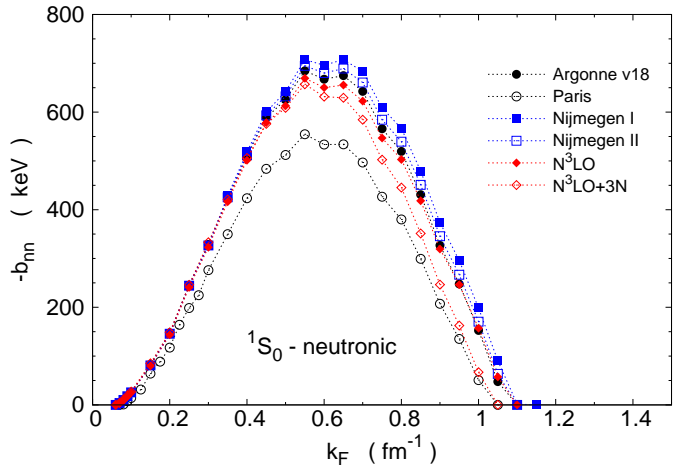}}
\caption{
 \emph{In-medium} di-neutron binding energy in pure neutron matter 
       as function of the Fermi momentum $k_F$.
  Symbols follow the same convention as in Fig.~\ref{meff_snm}.
}
\label{bnn1s0n}       
\end{figure}

Results for di-neutrons in pure neutron matter are shown 
in Fig.~\ref{bnn1s0n}, where we plot $b_{nn}$ as a function 
of $k_F$ considering all six interactions in this study,
applying the same notation as in the previous case.
Here we also use keV units for the energy scale.
Note that all interactions display similar behavior over $k_F$,
with appearance of di-neutrons at $k_F$ above 0.06~fm$^{-1}$
and disappearance at $k_F\approx 1.05$~fm$^{-1}$ in the case of
Paris and N$^{3}$LO+3N interactions, 
and at $k_F\approx 1.1$~fm$^{-1}$ for the rest.
The maximum binding takes place in the vicinity of
$k_F\approx 0.6$~fm$^{-1}$, with Paris potential 
leading to the lowest binding of $\sim\!-$550~MeV.
Note that the behavior of $b_{nn}$ in this case shows some
quantitative resemblance to that found for the $^{1}\textrm{S}_{0}$
channel in SNM (c.f. Fig.~\ref{bnn1s0snm}).

\subsection{Hugenholtz-van Hove theorem}
\label{hvh}
The Hugenholtz-van Hove (HvH) theorem~\cite{Hugenholtz1958}
states a very general result that relates the mean energy of a bound system,
$E/N$, with its chemical potential $\mu$.
At zero temperature
this relationship establishes that~\cite{Baldo99}
\begin{equation}
\label{pressure}
p = -\frac{E}{V} + \frac{N}{V} \mu \;,
\end{equation}
where $p$ represents the pressure, $E/V$ the energy density, 
$N/V$ the particle density, and $\mu$ the chemical potential.
The latter should be extracted from the derivative of energy with 
respect to the number of particles.
In the BHF approximation at zero temperature 
the chemical potential coincides with the 
Fermi energy $e_F$ of the system
--given by the sp energy at the Fermi momentum $k_F$-- 
hence relying on the auxiliary potential $U(k)$. 
Since at saturation the pressure vanishes,
Eq.~(\ref{pressure}) reduces to
\begin{equation}
\label{hvhtheorem}
\frac{E}{A} = e_F\,,
\end{equation}
with $A$ the nucleon number.
It has been known for some time that the HvH theorem is not 
satisfied in the BHF approximation for nuclear matter at zero temperature.
This limitation has led to go beyond BHF by including higher order
contributions in the hole expansion for the nucleon 
self-energy~\cite{Czerski2002,Zuo1999,Grange1987}.
Such an extension goes beyond the scope of this work. 
However, it is still instructive to assess the extent to which 
HvH theorem is violated within the BHF approach for the 
interactions considered in this work.

In Table~\ref{tab:1} we list the internucleon potentials
together with their respective Fermi momentum 
and density $\rho$ at saturation.
The sixth column displays the difference between the mean energy
$E/A$ and the Fermi energy, $e_F=e(k_F)$. 
If HvH was satisfied at the saturation point, 
then Eq.~\ref{hvhtheorem} would imply only zeros for this column.
Such is not the case, as we observe that the difference $(E/A-e_F)$ 
is comparable to $-\!E/A$.
The weakest violation of HvH theorem 
occurs for N$^{3}$LO+3N chiral interaction, where $E/A-e_F=8.2$~MeV.
\begin{table} [ht]
\caption{HvH theorem check at saturation point}
\label{tab:1}       
\begin{tabular}{l ccccc}
\hline
\noalign{\smallskip}
Potential & $k_F$      & $\rho$  
& $\textstyle{\frac{E}{A}}$ 
& $e_F$ 
& $\textstyle{\frac{E}{A}}-e_F$ \\
\noalign{\smallskip}
          &  [fm$^{-1}$]&[fm$^{-3}$] & [MeV] & [MeV] &   [MeV]   \\
\noalign{\smallskip}
\hline
\noalign{\smallskip}
AV18          & 1.52 & 0.237 & -16.8 & -34.5 & 17.7\\
Paris         & 1.53 & 0.242 & -16.3 & -33.7 & 17.4\\
Nijmegen I    & 1.68 & 0.320 &-20.6 & -40.5 & 19.9\\
Nijmegen II   & 1.63 & 0.293 &-18.3 & -36.8 & 18.5\\
N$^3$LO       & 1.85 & 0.428 & -25.7 & -45.9 & 20.2\\
N$^3$LO+3N    & 1.30 & 0.148 &-12.1 & -20.3 &  8.2\\
\noalign{\smallskip}\hline
\end{tabular}
\end{table}

The inclusion of higher-order correlations to mitigate the
violation of the HvH theorem has been investigated in
Refs.~\cite{Czerski2002,Zuo1999,Grange1987}.
In the present context, this extension would require a significant
amount of work.
Implications of these considerations in actual calculations
remain to be seen, particularly regarding the coexistence of
sp solutions in the case of SNM.

\section{Summary and conclusions}
\label{summary}
We have investigated the role of di-nucleon bound states in
homogeneous nuclear matter in the cases of isospin-symmetric 
matter and pure neutron matter.
The study has been based on the BHF approach at zero temperature,
considering modern bare internucleon interactions,
including the case of 3NFs based on chiral N$^{3}$LO with
three-nucleon forces up to N$^{2}$LO.
Special attention is paid to the occurrence of di-nucleon 
bound states structures in the $^1\textrm{S}_0$ and $^3\textrm{SD}_1$
channels, whose explicit treatment is critical for the stability 
of self-consistent solutions at sub-saturation densities.
An analysis of these solutions is made by comparing their
associated energy per nucleon $E/A$, effective masses
and \emph{in-medium} di-nucleon binding energies.

An important result from this work is that coexistence of
sp solutions in SNM withing the BHF approximation,
occurring at Fermi momenta in the range $0.13-0.3$~fm$^{-1}$
and reported in Ref.~\cite{Arellano15},
is a robust property of the system which does not depend on
the choice of realistic internucleon potentials.
Additionally, all interactions yield very similar behavior
of $E/A$ at fermi momenta $k_F\lesssim$~fm$^{-1}$.
At higher densities the interactions exhibit their differences,
resulting in different saturation points in the case of SNM,
or different growth of $E/A$ (i..e. pressure) as function of
the density.
Additionally, we also obtain effective masses larger than bare
masses at sub-saturation densities, feature shared by all 
interactions. 
In the case of SNM, effective masses in phase I can reach
up to four times the bare mass, while in the case of phase II
a maximum ratio of $m^*/m\approx 1.5$ is found 
at $k_F\approx 0.5$~fm$^{-1}$.
In the case of pure neutron matter, the highest effective masses
occur in the range $0.25\lesssim k_F\lesssim 0.5$~fm$^{-1}$,
where $m^*/m$ can reach up to $\sim\! 1.2$.

Di-nucleons have also been investigated in both SNM and neutronic 
matter, identified from singularities in the $g$ matrix at starting 
energies below particle-particle threshold energy. 
In this work we obtain results consistent to those reported 
in Ref.~\cite{Arellano15}, but not restricted anymore to AV18 in SNM.
Bound states are identified at sub-saturation densities,
with deuterons in phase I bound at energies nearly twice that 
in free space.
Deuterons in phase II reach maximum binding at 
$k_F\approx 0.7$~fm$^{-1}$, with binding energies comparable 
to that in free space.
These \emph{in-medium} bound states get dissolved for
$k_F\gtrsim 1.3$~fm$^{-1}$, for all the interactions considered.
Di-nucleons in channel $^1\textrm{S}_0$ are found in both,
SNM and neutronic matter. Their binding is much weaker to that
for deuterons, reaching deepest values between $-\!700$ and
$-\!500$~keV. 
In this particular channel di-nucleons get dissolved at Fermi
momenta above 1.1~MeV, feature shared by all interactions considered.
Overall, the binding properties of di-nucleons appear quite
comparable, pointing also to their robustness under the interaction
being considered.

The occurrence of di-nucleons in nuclear matter is
closely related to nuclear pairing phenomena, mechanism
responsible for the formation of Cooper pairs and the
emergence of superfluid and superconducting states of matter 
\cite{Broglia13,Sedrakian06}.
This aspect, addressed to some extent in Ref.~\cite{Arellano15},
has been omitted here since there would be no substantial new 
information.
On this regard we have checked that all interactions behave very 
similar to AV18 potential.
At a more basic level, in this work we have investigated the 
degree of fulfillment of Hugenholtz-van Hove theorem,
finding that BHF approach alone fails considerably. 
However, studies reported in 
Refs.~\cite{Czerski2002,Zuo1999,Grange1987}.
point that inclusion of higher order terms in the
series expansion would remedy this limitation. 
Efforts to include rearrangement corrections are underway.

\begin{acknowledgement}
F.I. thanks CONICYT fellowship Beca Nacional, Contract No. 221320081.
This work was supported in part by STFC through Grants ST/I005528/1,
ST/J005743/1 and ST/L005816/1.
Partial support comes from ``NewCompStar'', COST Action MP1304.
\end{acknowledgement}

%

\begin{thebibliography}{29}

\bibitem{Baldo99}
M.~Baldo, ed., \emph{Nuclear Methods and the Nuclear Equation of State}, Vol.~8
  of \emph{International Review of Nuclear Physics} (World Scientific,
  Singapore, 1999).

\bibitem{Entem2003}
D.R. Entem, R.~Machleidt, Phys. Rev. C \textbf{68}, 041001 (2003).

\bibitem{Dickhoff2008}
W.H. Dickhoff, D.~Van~Neck, \emph{Many-Body Theory Exposed} (World Scientific,
  Singapore, 2008).

\bibitem{Dewulf03}
Y.~Dewulf, W.H. Dickhoff, D.~Van~Neck, E.R. Stoddard, M.~Waroquier, Phys. Rev.
  Lett. \textbf{90}, 152501 (2003).

\bibitem{Song98}
H.Q. Song, M.~Baldo, G.~Giansiracusa, U.~Lombardo, Phys. Rev. Lett.
  \textbf{81}, 1584 (1998).

\bibitem{Amos00}
K.~Amos, P.J. Dortmans, H.V. von Geramb, S.~Karataglidis, J.~Raynal,
  \emph{Advances in Nuclear Physics}, Vol.~25 of \emph{Advances in Nuclear
  Physics} (Springer, New York, 2000).

\bibitem{Arellano15}
H.F. Arellano, J.P. Delaroche, Eur. Phys. Journal A \textbf{51}, 1 (2015).

\bibitem{Wir95}
R.B. Wiringa, V.G.J. Stoks, R.~Schiavilla, Phys. Rev. C \textbf{51}, 38 (1995).

\bibitem{Hugenholtz1958}
N.M. Hugenholtz, L.~van Hove, Physica \textbf{24}, 363 (1958).

\bibitem{Baldo00}
M.~Baldo, A.~Fiasconaro, Phys. Lett. B \textbf{491}, 240 (2000).

\bibitem{Paris}
M.~Lacombe, B.~Loiseau, J.M. Richard, R.V. Mau, J.~C\^ot\'e, P.~Pir\`es,
  R.~de~Tourreil, Phys. Rev. C \textbf{21}, 861 (1980).

\bibitem{Nijmegen}
V.G.J. Stoks, R.A.M. Klomp, C.P.F. Terheggen, J.J. de~Swart, Phys. Rev. C
  \textbf{49}, 2950 (1994).

\bibitem{Holt2010}
J.W. Holt, N.~Kaiser, W.~Weise, Phys. Rev. C \textbf{81}, 024002 (2010).

\bibitem{Hebeler2010a}
K.~Hebeler, A.~Schwenk, Phys. Rev. C \textbf{82}, 014314 (2010).

\bibitem{Carbone2014}
A.~Carbone, A.~Rios, A.~Polls, Phys. Rev. C \textbf{90}, 054322 (2014).

\bibitem{Nogga2006}
A.~Nogga, P.~Navr\'atil, B.R. Barrett, J.P. Vary, Phys. Rev. C \textbf{73},
  064002 (2006).

\bibitem{omponline}
H.F. Arellano, \emph{omp-online website} (2006), $U(k;k_F)$ solutions
  accessible for download, \texttt{\url{http://www.omp-online.cl}}.

\bibitem{Carbone2013}
A.~Carbone, A.~Polls, A.~Rios, Phys. Rev. C \textbf{88}, 044302 (2013).

\bibitem{Tews2016}
I.~Tews, S.~Gandolfi, A.~Gezerlis, A.~Schwenk, Phys. Rev. C \textbf{93}, 024305
  (2016).

\bibitem{Sammarruca2015}
F.~Sammarruca, L.~Coraggio, J.W. Holt, N.~Itaco, R.~Machleidt, L.E. Marcucci,
  Phys. Rev. C \textbf{91}, 054311 (2015).

\bibitem{Lombardo2006}
Z.H. Li, U.~Lombardo, H.J. Schulze, W.~Zuo, L.W. Chen, H.R. Ma, Phys. Rev. C
  \textbf{74}, 047304 (2006).

\bibitem{Chamel2013}
N.~Chamel, Phys. Rev. Lett. \textbf{110}, 011101 (2013).

\bibitem{Baldo2014}
M.~Baldo, G.F. Burgio, H.J. Schulze, G.~Taranto, Phys. Rev. C \textbf{89},
  048801 (2014).

\bibitem{Isaule2016}
F.~Isaule, H.F. Arellano, A.~Rios, Phys. Rev. C \textbf{94}, 044317 (2016).

\bibitem{Czerski2002}
P.~Czerski, A.~De~Pace, A.~Molinari, Phys. Rev. C \textbf{65}, 044317 (2002).

\bibitem{Zuo1999}
W.~Zuo, I.~Bombaci, U.~Lombardo, Phys. Rev. C \textbf{60}, 024605 (1999).

\bibitem{Grange1987}
P.~Grange, J.~Cugnon, A.~Lejeune, Nuclear Physics A \textbf{473}, 365  (1987).

\bibitem{Broglia13}
R.A. Broglia, V.~Zelevinsky, eds., \emph{Fifty Years of Nuclear BCS} (World
  Scientific, Singapore, 2013).

\bibitem{Sedrakian06}
A.~Sedrakian, J.W. Clark, M.~Alford, eds., \emph{Pairing in Fermionic Systems}
  (World Scientific, Singapore, 2006).

\end{thebibliography}
%

\end{document}